\documentclass[aps,reprint,twocolumn,showpacs,preprintnumbers,amsmath,nofootinbib,amssymb,superscriptaddress,showkeys]{revtex4-1}

\usepackage{epsfig}
\usepackage{slashed}

\begin{document}

\title{The parity-violating nucleon-nucleon force in the $1/N_c$ expansion}
\author{Daniel R. Phillips}

\affiliation{Institute of Nuclear and Particle Physics and
Department of Physics and Astronomy, Ohio University, Athens, Ohio 45701, USA}

\author{Daris Samart}

\affiliation{Department of Applied Physics, Faculty of Sciences and Liberal
Arts, Rajamangala University of Technology Isan, Nakhon Ratchasima, 30000,
Thailand}
\affiliation{Thailand Center of Excellence in Physics (ThEP), Commission on Higher Education, Bangkok 10400, Thailand}

\author{Carlos Schat}
\affiliation{Departamento de F\'{\i}sica, FCEyN, Universidad de Buenos
Aires and IFIBA, CONICET,
Ciudad Universitaria, Pab.~1, (1428) Buenos Aires, Argentina}

\begin{abstract} 
Several experimental investigations have observed parity violation in
nuclear systems---a consequence of the weak force between quarks.  We apply the $1/N_c$ expansion of QCD to the
P-violating T-conserving component of the nucleon-nucleon (NN) potential. We show 
there are two leading-order operators, both of which affect $\vec{p}p$ scattering at order
$N_c$. We find an additional four operators at order  $N_c^0 \sin^2 \theta_W$
and six  at  ${\cal O}(1/N_c)$. Pion exchange in the PV NN force is
suppressed by $1/N_c$ and $\sin^2 \theta_W$, providing a quantitative explanation for its non-observation up to this time. 
The large-$N_c$ hierarchy of other PV NN force mechanisms is  consistent with estimates of the couplings in
phenomenological models. The PV observed in $\vec{p}p$ scattering data is compatible with
natural values for the strong and weak coupling constants: there is no evidence of fine
tuning. 

\end{abstract}

\maketitle

The strong-nuclear and electromagnetic forces play the most prominent role
in proton-proton ($pp$) scattering. There are also parity-violating (PV) $pp$
interactions, which manifest the presence of weak interactions between the quarks inside
each proton.  Measurements of  longitudinal beam asymmetries $\sim 10^{-7}$ at Bonn~\cite{Eversheim},
PSI~\cite{PSI}, and TRIUMF~\cite{Berdoz} demonstrate that PV
nucleon-nucleon (NN) forces exist. PV in NN systems is also probed via an asymmetry in the
reaction $\vec{n}p \rightarrow d \gamma$~\cite{Gericke:2011zz,Fomin:2013ssa}. And {\it ab initio} calculations of few-nucleon systems allow
us to take models of the PV NN force and predict, e.g., the longitudinal asymmetry in
${}^3$He$(\vec{n},p)^3$H~\cite{Viviani:2010qt}, which is soon to be measured~\cite{Bowmanproposal}. Nuclear parity violation is also observed in,
e.g., the radiative decay of the first excited state of ${}^{19}$F, but there 
theoretical uncertainties in the relationship between the observable and the model of the
PV NN force are harder to quantify.  Much work has gone into
constraining the PV NN force from a variety of nuclear experiments, see
Refs.~\cite{Schindler:2013yua,Haxton:2013aca} for recent reviews.

The prevailing paradigm in such analyses is based on single-meson exchange between
nucleons, most commonly in the framework developed by Desplanques, Donoghue, and Holstein
(DDH)~\cite{Desplanques:1979hn}. The quantum numbers of the exchanged mesons determine the
operator structures that contribute, while operator coefficients involve products of
strong and weak meson-nucleon-nucleon coupling constants. In this paper we show that
Standard Model (SM) couplings and the $1/N_c$ expansion of
QCD predict the operators, and the sizes of the associated coefficients,
which appear in the PV NN potential. 

An alternative framework---suitable for studying PV at very low energies---that systematizes
pioneering studies~\cite{Danilov,DesplanquesMissimer} has recently
emerged~\cite{Phillips:2009,Schindler:2009,Griesshammer:2011}, but has, as yet, been
applied to far fewer experiments. The extension of chiral perturbation theory to
few-nucleon systems, $\chi$EFT~\cite{Epelbaum:2008ga} has also 
been invoked~\cite{Zhu:2003,Liu:2007,Viviani:2014zha,deVries:2013fxa,deVries:2014vqa}.
In $\chi$EFT the one-pion-exchange piece of the
PV NN force dominates, with all other effects suppressed by two orders in the
chiral expansion.  

One-pion exchange gives the long-distance parity-conserving potential, and drives many of the properties of light nuclei. 
But, thus far, experimental data show no evidence for pion exchange in the PV NN force: only 
upper bounds on its impact on observables have been obtained. 
We will show that the smallness of the PV NN operator associated
with one-pion exchange is a consequence of the large-$N_c$ expansion. 

Originally suggested by 't Hooft~\cite{'tHooft:1973jz}, this technique notes
that QCD has a ``hidden", perhaps small, parameter in $1/N_c$. Multiple simplifications of QCD occur in the limit $N_c \rightarrow \infty$. In particular,  the expansion
in powers of $1/N_c$ provides insights about baryons~\cite{Jenkins:1998wy,Manohar:1998xv}. In the context of nuclear forces
the $1/N_c$ expansion was  used to study the NN potential~\cite{KS96,Kaplan:1996rk}. These
works analyzed the NN potential for momenta of order $N_c^0$, i.e., $p \sim
\Lambda_{QCD}$, and found that it is an expansion in $1/N_c^2$. This hierarchy between different contributions to the
NN potential is roughly borne out in the Nijm93~\cite{St94} NN potential. This analysis was extended to the 3N potential~\cite{Phillips:2013rsa};
here we tackle the PV component of the NN force. Some of
our results have been obtained within the chiral soliton
model~\cite{Kaiser:1989fd,Meissner:1998pu,Grach:1988bv}, 
or from consistency relations for PV pion-nucleon 
scattering~\cite{Zhu:2009nj}. But a model-independent derivation of all pertinent scalings and comparison with
experimental data and phenomenological potentials appears here for the first
time. 

The fact that  $\sin^2\theta_W \approx 0.23$~\cite{PDG} is key to  the
hierarchy of PV NN operators.  The SM effective Lagrangian for the
PV four-quark operators involving $u$ and $d$ quarks \cite{Donoghue:1992dd} is:
\begin{eqnarray}
\mathcal{L}_{4q}^{\rm eff} &=& - \frac{G_F}{\sqrt 2}\,\cos^2\theta_C \sum_{a=1,2}\Big( V^{a}_\mu - A^{a}_\mu\Big)^2
\nonumber\\
&\;& -\,  \frac{G_F}{\sqrt 2}\,\Big( \cos2\theta_W\,V^3_\mu - A^3_\mu - 2\,\sin^2\theta_W\,I_\mu \Big)^2
\nonumber\\
&=& {\sqrt 2}\,{G_F}\,\Big\{\cos^2\theta_C \sum_{a=1,2} \,V_\mu^a\,A^{\mu a}
+ \cos2\theta_W\,V^3_\mu\,A^{\mu 3} \nonumber\\
&&- 2\,\sin^2\theta_W\,I_\mu\,A^{\mu 3}  \Big\}
 +\, \cdots\,,
 \label{eq:SMquarkoperators}
\end{eqnarray}
where we kept only the PV terms. Here $G_F=1.16 \times 10^{-5} \
{\rm GeV}^{-2}$, $\cos^2\theta_C = 0.946$ and we dropped Cabibbo suppressed terms. 
The currents are 
$V_\mu^a=\frac{1}{2}\,\bar q\,\gamma_\mu \tau^a q$, 
$A_\mu^a=\frac{1}{2}\,\bar q\,\gamma_\mu \gamma^5 \tau^a q$ and 
$I_\mu=\frac{1}{6}\,\bar q\,\gamma_\mu q$ 
respectively.  Importantly, the factor $\sin^2\theta_W$\, accompanies  the
product of isoscalar and axial currents, which is the only source of
$\Delta I=1$ operators. $\Delta I=0$ and $\Delta I=2$ operators have pre-factors of
${\cal O}(1)$. Running from the $Z^0$-mass down to the strong scale $\Lambda_\chi \sim$ 1 GeV does
not significantly modify this hierarchy~\cite{Dai:1991bx,Tiburzi:2012hx}.

Now we estimate the NN matrix elements of quark
operators in Eq.~(\ref{eq:SMquarkoperators}) using the
Hartree expansion for the nuclear Hamiltonian in the large-$N_c$
limit~\cite{Dashen:1994qi,Kaplan:1996rk}
\begin{eqnarray}
H = N_c \sum_{s, t, u} v_{stu} \left(\frac{S}{N_c}\right)^s  
\left(\frac{I}{N_c}\right)^t \left(\frac{G}{N_c}\right)^u \ .
\label{eq:Hartree}
\end{eqnarray}
The explicit factors of $1/N_c$ ensure that an $m$-body interaction scales
 as $1/N_c^{m-1}$~\cite{Witten:1979kh}. The coefficients are ${\cal O}(1)$ functions of the
momenta. We take a quark basis  for the operators:
\begin{equation} 
S^i = q^\dagger  \frac{\sigma^i}{2} q , \quad   
I^a = q^\dagger  \frac{\tau^a}{2} q , \quad  
G^{ia} = q^\dagger  \frac{\sigma^i \tau^a}{4} q \ ,
\label{qop}
\end{equation}
which generate an $SU(4)$ algebra.  We wish to
take their matrix elements in the $|NN \rangle$ piece of the
hadronic Hilbert space.  $S$, $I$, $G$ in Eq.~(\ref{eq:Hartree})  can have any
nucleon index; we denote by $O_{\alpha}$ the
nucleon ($\alpha= 1,2$) on which they act. Products of operators acting on the same nucleon 
are reduced to a single operator using relations for  powers of  $S$, $I$, $G$~\cite{Dashen:1994qi,Phillips:2013rsa}. Matrix elements of $S$ and $I$ between
nucleon states are ${\cal O}(1)$, while matrix elements of $G$ are ${\cal O}(N_c)$. 
The mass of the nucleon, $m_N$, scales as $N_c$. This implies that any leading-order (LO) large-$N_c$ NN potential is
(modulo exchange diagrams) local: it is a function of the relative co-ordinate
${\bf r}$; or, equivalently, in momentum space, depends solely on the
difference of final- and initial-state relative momenta, ${\bf p}_- \equiv {\bf
p}' - {\bf p}$. The combination ${\bf p}_+  \equiv {\bf p}' + {\bf p}$ can
appear only via relativistic corrections, and so its occurrence is always
suppressed by a factor of $1/N_c$. Both ${\bf p}_-$ and ${\bf p}_+$ are
parity odd, with ${\bf p}_-$ (${\bf p}_+$) being even (odd) under time
reversal. 

We now use these momentum operators to counterbalance the spin-flavor
structures obtained after using the reduction rules in Eq.~(\ref{eq:Hartree}).
We do this to obtain a Hamiltonian that is a rotational scalar,
time-reversal even, and parity odd.  As to its isospin transformation
properties, we have already seen that $\Delta I=0, 1, 2$ operators arise in the
SM. At the hadronic level the leading-in-$N_c$ operators are:
\begin{eqnarray}
U_{PV}^{N_c} = N_c\,\Big(U_P^1(\mathbf{p}_-^2)\,\big[ \mathbf{p}_-\cdot\,(\sigma_1\times
\sigma_2)\,\tau_1\cdot\,\tau_2 \big]\nonumber\\
+ U_P^2(\mathbf{p}_-^2)\,\big[ \mathbf{p}_-\cdot\,(\sigma_1\times
\sigma_2)\,[\tau_1\,\tau_2]_{_2}^{zz} \big] \Big)\,,
\label{eq:LO}
\end{eqnarray}
where $[\dots]_{_2}$ denotes a symmetric and traceless rank-two tensor.
These mediate $\Delta I=0,2$ transitions.   Since ${\bf p}_-$ is ${\cal O}(1)$ an
arbitrary function of ${\bf p}_-^2$ can appear as a pre-factor without changing
the $N_c$ order of any contribution.  

The four ${\cal O }(N_c^0 \sin^2 \theta_W)$  operators---all $\Delta I=1$---are:
\begin{widetext}
\begin{eqnarray}
U_{PV}^{N_c^0} &=& N_c^0\,\Big(U_P^3(\mathbf{p}_-^2)\,\big[ \mathbf{p}_+\cdot\,(\sigma_1\,\tau_1^z - \sigma_2\,\tau_2^z) \big]
+ U_P^4(\mathbf{p}_-^2)\,\big[ \mathbf{p}_-\cdot\,(\sigma_1 + \sigma_2)\,(\tau_1\times\,\tau_2)^z \big]
\nonumber\\
&&\quad +\, U_P^5(\mathbf{p}_-^2)\,\big[ \mathbf{p}_-\cdot\,(\sigma_1 \times\, \sigma_2)\,(\tau_1 + \tau_2)^z \big]
 +\, U_D^1(\mathbf{p}_-^2)\,\big[
[(\mathbf{p}_+\times\,\mathbf{p}_-)^i\,\mathbf{p}_-^j]_{_2}\cdot\,[\sigma_1^i\,
\sigma_2^j]_{_2}\,(\tau_1\times\,\tau_2)^{z} \big] \Big)\,.
\label{eq:NLO}
\end{eqnarray}
At ${\cal O}(1/N_c)$ there are a number of additional $\Delta I=0,2$ operators that arise:
\begin{eqnarray}
&&U_{PV}^{N_c^{-1}} = N_c^{-1}\,\Big(  U_P^6(\mathbf{p}_-^2)\,\big[ \mathbf{p}_-\cdot\,(\sigma_1\times \sigma_2) \big]
+ U_P^7(\mathbf{p}_-^2)\,\big[ \mathbf{p}_+^2\,\mathbf{p}_-\cdot\,(\sigma_1\times \sigma_2)\,\tau_1\cdot\,\tau_2\big]
+\, U_P^8(\mathbf{p}_-^2)\,\big[ \mathbf{p}_+\cdot\,(\sigma_1 - \sigma_2) \big]\nonumber\\
&& \;
+ U_P^9(\mathbf{p}_-^2)\,\big[ \mathbf{p}_+\cdot\,(\sigma_1 - \sigma_2)\,\tau_1\cdot\,\tau_2 \big]
+\, U_P^{10}(\mathbf{p}_-^2)\,\big[ \mathbf{p}_+\cdot\,(\sigma_1 -
\sigma_2)\,[\tau_1\,\tau_2]_{_2}^{zz} \big]
+\, U_P^{11}(\mathbf{p}_-^2)\,\big[ \mathbf{p}_+^2\,\mathbf{p}_-\cdot\,(\sigma_1\times
\sigma_2)\,[\tau_1\,\tau_2]_{_2}^{zz} \big]\Big),
\label{eq:NLO2}
\end{eqnarray}
\end{widetext}
while at ${\cal O}(1/N_c^2)$ corrections to the coefficient functions in Eq.~(\ref{eq:NLO}) and additional $\Delta I=1$ operators appear.
Note that the $1/N_c$ expansion says
very little about the coefficient functions $U_P^{1}$--$U_P^{11}$ and $U_{D}^1$; the only constraint on
them is that they should be ${\cal O}(1)$. 

Within each isospin sector the expansion is thus in $1/N_c^2$, as for the
strong NN and NNN force. Since $\Delta I=1$ operators are 
suppressed by $\sin^2 \theta_W$ and $1/N_c$ the two
operators in Eq.~(\ref{eq:LO}) give the entire PV NN force up to corrections that are
formally of relative order $1/N_c^2$, $\sin^2 \theta_W/N_c$. Below we will argue,
though, that the numerical suppression is not quite the $\approx 10$\% this implies.

An expansion in momenta would reduce Eqs.~(\ref{eq:LO})--(\ref{eq:NLO2}) to the five
operators that describe the S-P transitions
~\cite{Danilov,Girlanda:2008ts,Phillips:2009}. Here we do not take the low-energy 
limit as we want to compare with the full DDH potential.
Furthermore, at the 221 MeV energy of the TRIUMF $\vec{p}p$ experiment that we
also seek to describe, an expansion in powers of momenta is
not trustworthy.  

We now compare our result to the PV NN potential
of DDH. The relevant expressions can be found in  \cite{Desplanques:1979hn} and
\cite{Haxton:2013aca}. The one-meson-exchange diagrams from the
weak and strong meson-nucleon Hamiltonians given there yield the DDH potential 
as a set of operators, each of which is multiplied by one strong and
one weak meson-nucleon-nucleon coupling. Up to ${\cal O}(N_c^0 \sin^2 \theta_W)$ only one
spin-flavor structure is produced by the $1/N_c$ analysis that does not appear in the
DDH potential. It is the operator multiplied by the coefficient function
$U_D^1$ and corresponds to a tensor constructed from ${\bf L}$ and ${\bf
p}_-$ coupled to the rank-two spin tensor. This structure is not generated
straightforwardly in the meson-exchange picture.

The rest of the DDH structures can each be matched to one structure in the LO
or ${\cal O}(N_c^0 \sin^2 \theta_W), {\cal O}(1/N_c)$ potentials. DDH made a prediction for the
strength of these operators based on standard values for the strong
meson-nucleon-nucleon couplings and estimates of the ``best values" for the
weak couplings. We will use these weak-coupling estimates as
our point of comparison (but see
Ref.~\cite{AdelbergerHaxton}). In order to extract values for the weak couplings
from our $1/N_c$ analysis, we recall the $N_c$-scaling rule of the strong
couplings from Ref.~\cite{Kaplan:1996rk}:
\begin{equation} 
g_\omega \sim \sqrt{N_c}\,, \; \;  
g_\rho \sim \frac{1}{\sqrt{N_c}} \,,\;\; 
\xi_V \sim N_c \,, \;\; 
\xi_S \sim \frac{1}{N_c}.
\label{eq:strongNc}
\end{equation}
We count the pion's coupling as $\bar g_{\pi NN} \sim
\sqrt{N_c}$.
This, together with the Goldberger-Treiman relation, means that the pseudoscalar
$\pi$NN coupling which appears in the DDH potential,
$g_{\pi NN} =\frac{m_N}{\Lambda_\chi} \bar g_{\pi NN}$,
scales as $N_c^{3/2}$. In a similar vein, we replaced DDH's
parameters $\chi_{V,S}$ by $m_N \xi_{V,S}/\Lambda_\chi$ and $h_\rho^{1'}$ by $m_N
h_\rho^{1'}/\Lambda_\chi$,  so that there are no spurious factors of $N_c$ appearing in the
coefficients of operators via the nucleon mass.   $\Lambda_\chi \sim 1 \ {\rm GeV}  $ suppresses higher dimensional operators and is independent
of $N_c$.

We then extract the $N_c$ and
$\sin^2\theta_W$ scalings of the weak couplings in DDH potential as:
\begin{eqnarray}
&& h_\rho^0 \sim \sqrt{N_c}\,, \; \; h_\rho^2 \sim \sqrt{N_c}\,,
\nonumber\\
&& \frac{h_\rho^{1'}}{\sin^2\theta_W} \lesssim \sqrt{N_c}\,,\; \;  
\frac{h_\omega^1}{\sin^2\theta_W} \sim \sqrt{N_c}\,, \; \;\nonumber\\
&&   \frac{h_\rho^1}{\sin^2\theta_W} \lesssim \frac{1}{\sqrt{N_c}}\,, \; \; 
\frac{h_\pi^1}{\sin^2\theta_W} \lesssim \frac{1}{\sqrt{N_c}}\,,
\; \; h_\omega^0 \sim \frac{1}{\sqrt{N_c}}\, .
\label{eq:weakNc}
\end{eqnarray}
Since they arise from the
$I_\mu\,A^{\mu 3}$ product in the effective four-quark Lagrangian, the
$\Delta I=1$ couplings must all include a
factor of $\sin^2 \theta_W$. The bounds on the scalings of $h_\rho^{1}$, $h_\rho^{1'}$, $h_\pi^{1}$ follow
from requiring that the large-$N_c$ scaling is not violated,
while the scalings of $ h_\rho^{0,2},h_\omega^{0,1} $ are needed in order that
the $U's $ in Eqs.~(\ref{eq:LO},\ref{eq:NLO},\ref{eq:NLO2}) scale as $ {\cal
O}(N_c^0)$. 
Some of these results in Eq.~(\ref{eq:weakNc}) were previously derived in
 soliton models~\cite{Kaiser:1989fd,Meissner:1998pu,Grach:1988bv}.  And Ref.~\cite{Zhu:2009nj} computed the large-$N_c$ scaling of $h_\pi^1$, but did not 
account for its $\sin^2 \theta_W$ suppression.

The two operators in Eq.~(\ref{eq:LO}) give the entire LO PV NN potential. 
When written in terms of the DDH couplings they are proportional to
$g_\rho h_\rho^{0,2} \chi_V/m_N$. Taking the product of all scalings 
in this expression verifies that the potential in Eq.~(\ref{eq:LO}) is  ${\cal O}(N_c)$, but also shows that one of the factors of $N_c$ is 
associated with the factor of $m_N/\Lambda_\chi \sim 1$ that (implicitly) occurs in the DDH coupling $\chi_V$. 
This effectively demotes the LO piece of the PV NN potential to a numerical size typical of 
a ${\cal O}(N_c^0)$ contribution. We therefore 
conclude that the two operators in Eq.~(\ref{eq:LO}) {\it determine the
parity-violating NN force up to $\approx$ 30\% corrections.}

\begin{figure}
\begin{center}
\includegraphics[width=7cm,clip=true]{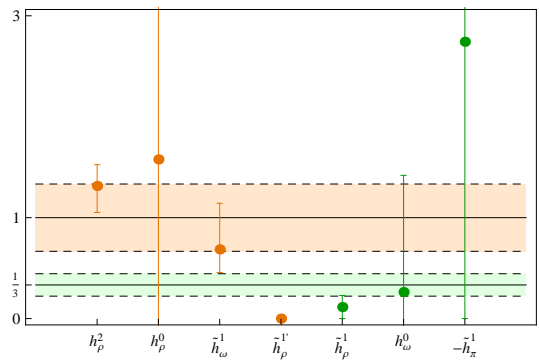}
\end{center}
\vspace*{-2mm}
\caption{(Color online) The hierarchy of weak couplings for the DDH best values
\cite{Desplanques:1979hn}, re-scaled by the average of $h^2_\rho$ and $\tilde h^1_\omega$.
The bands show the large-$N_c$ predictions, with their associated ${\cal O}(1/N_c)$
error bars. The tildes on couplings indicate that both the
$N_c$-prediction and the DDH best value have been divided by $\sin^2 \theta_W$. The orange (green) points are couplings which are ${\cal O}(\sqrt{N_c})$ (${\cal O}(1/\sqrt{N_c})$). Error bars indicate the DDH ``reasonable ranges".}
\label{fig-DDHplot}
\end{figure}

Although the DDH ranges are large, the preferred values fall in the
relatively narrow bands predicted by the $1/N_c$ hierarchy, except for $
h^{1'}_\rho$ and $h^1_\pi$, see Fig.~\ref{fig-DDHplot}. Notably, the two
LO operators are associated with the largest couplings, $h^0_\rho$, $h^2_\rho$.  The DDH best value for the
coupling $h^1_\omega$ is also within 30\% of the natural value once the $\sin^2
\theta_W$ suppression is taken into account.  The $\sin^2
\theta_W/N_c$ suppressed couplings include $h_\pi^1$. This is in contrast to DDH, who have a $h_\pi^1$
glaringly
larger than the large-$N_c$ prediction.  A much smaller $h_\pi^1$ is found in
soliton models~\cite{Kaiser:1989fd,Meissner:1998pu}, and appears to 
be borne out by experiment (see, e.g. Fig.~3 
in Ref.~\cite{Haxton:2013aca}). Lastly, Ref.~\cite{Holstein:1981cg} used the quark model to argue that
the coupling $h_\rho^{1'}$ was small, and
as a consequence it has been neglected in many subsequent analyses. In
contrast, large-$N_c$ gives no reason that this coupling is any less important
than, say, $h_\pi^1$; both generate the operator structure $ (\sigma_1 +
\sigma_2)\,(\tau_1\times\,\tau_2)^z $ in Eq.~(\ref{eq:NLO}) and consistency
with the $N_c^0$ scaling of $U^4_P$ only requires that at least one of
$h_\rho^{1'}$, $h_\pi^1$ saturates the bounds given in
Eq.~(\ref{eq:weakNc}).

\begin{figure}
\vskip 5 mm
\begin{center}
\includegraphics[width=8cm,clip=true]{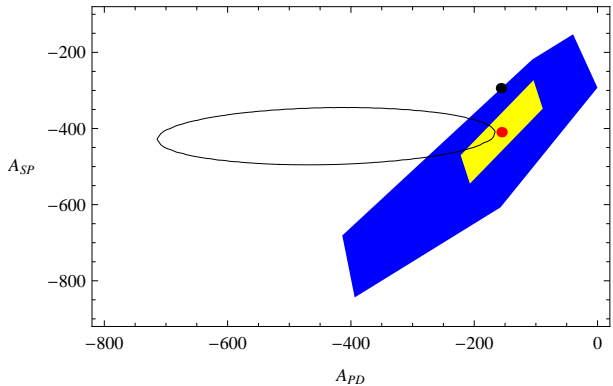}
\end{center}
\vspace*{-2mm}
\caption{(Color online) The ellipse gives the 90\% C.L. constraint from
experiment~\cite{Eversheim,PSI,Berdoz} on the combinations of couplings $A_{SP}$ and
$A_{PD}$ (in units of $10^{-7}$) via the analysis of
Refs.~\cite{Carlson:2001ma,Haxton:2013aca}.  
The black point corresponds to the DDH best value
and the red point is  obtained from naturalness and our large-$N_c$ analysis. 
The blue region is  found
by varying the weak and strong couplings by 30\% around their natural values.
The smaller yellow region is obtained by
only varying the weak couplings by 30\%.}
\label{fig-ellipses} 
\end{figure}

None of this, though, is a comparison at the level of observables. As already alluded
to, there are many problems with the extraction of weak
meson-nucleon-nucleon couplings from data, e.g. extracted weak
coupling constants depend on the strong coupling constants used~\cite{Haxton:2013aca}.
Constraining the products of weak and strong couplings from experiment may be a
better choice. Therefore we conclude our discussion
by considering the dominant combinations $g_\rho h_\rho^{0} , g_\rho h_\rho^{2},  g_\omega
h_\omega^{0} \sim {\cal O}(1)$ and $g_\omega h_\omega^{1} \sim {\cal O}(N_c \sin^2
\theta_W)$  asking what
they predict for experiments. All four contribute to $pp$ scattering. The $\vec{p}p$
asymmetry has been measured at 15~\cite{Eversheim}, 45~\cite{PSI}, and 221
MeV~\cite{Berdoz}. In the main, the first two experiments constrain the
PV-induced mixing between ${}^1$S$_0$- and ${}^3$P$_0$-waves, while the third
constrains  mixing between ${}^3$P$_2$- and $^1$D$_2$-waves. In plane-wave
Born approximation the  information can be parameterized by 
$A_{SP}$ and $A_{PD}$~\cite{Carlson:2001ma,Haxton:2013aca}, the pertinent
combinations of coupling constants governing these mixings. In the DDH
approach they are: 
\begin{eqnarray} 
A_{SP} &\equiv& g_\rho h_\rho^{pp} (2 + \chi_V) + g_\omega h_\omega^{pp} (2 + \chi_S), \nonumber\\ 
A_{PD} &\equiv& g_\rho h_\rho^{pp} \chi_V + g_\omega h_\omega^{pp} \chi_S,
\end{eqnarray} 
with $h_M^{pp}$ the combination of $\Delta I=0,1,2$ couplings
relevant for $pp$ scattering~\footnote{Note that two leading couplings $h_\rho^0$ and $h_\rho^2$
affect the $\vec{p}p$ asymmetry only via this combination. A lattice
QCD calculation of the iso-tensor piece of the PV NN interaction
would help break this degeneracy.}. The data from Refs.~\cite{Eversheim,PSI,Berdoz}
were analyzed in Ref.~\cite{Carlson:2001ma}, resulting in the constraints on
$A_{SP}$ and $A_{PD}$ shown in Fig.~\ref{fig-ellipses}. (See also the recent $\chi$EFT analysis~\cite{deVries:2013fxa,deVries:2014vqa}.)
While the variables for
the ellipse are motivated using plane-wave Born approximation, the calculation
is {\it not} done that way. It accounts for all initial- and final-state
(strong) $pp$ interactions, via a CD-Bonn
potential calculation of the corresponding wave functions~\cite{CDBonn}. 

To make a prediction for $A_{SP}$ and $A_{PD}$ we take $G_F f_\pi \Lambda_\chi \sim 1.0 \times 10^{-6}$ (with $f_\pi = 92.4 \ {\rm MeV} \sim \sqrt{N_c}$)
as the naturalness estimate for the LO weak couplings $h_\rho^{0,2}$.  Assuming a
natural value for $g_\omega \approx 4 \pi$~\cite{Furnstahl:1996wv}, we determine other
couplings by the $N_c$, $\sin^2 \theta_W$ scalings of Eqs.~(\ref{eq:strongNc})
and (\ref{eq:weakNc}), yielding $\{ -1.0 \, , -0.077 \, , -1.0 \, , -0.33 \, , -0.23 \}
\times 10^{-6} $ for the weak couplings $ \{ h_\rho^0, h_\rho^1, h_\rho^{2}, h_\omega^0,
h_\omega^1 \}$ and $ \{ 12. \, , 4.0 \, , -0.33 \, , 3.0 \} $  for the strong couplings $ \{
g_\omega, g_\rho, \xi_S, \xi_V \}$.  This is denoted by the red point in
Fig.~\ref{fig-ellipses}.  
All nine couplings should be assigned a 30\% error, due to omitted terms in the
$1/N_c$ expansion. Uncorrelated variation
over this range produces the blue shaded area in the figure. The yellow shaded area
is the result found from solely varying the five weak couplings. 
 The prediction for
$A_{SP}$ and $A_{PD}$ from large-$N_c$ and naturalness is thus consistent with the
constraints extracted in Ref.~\cite{Carlson:2001ma} within the combined
theoretical and experimental uncertainties. It shows no evidence of fine tuning. 
The black dot is obtained with DDH ``best values" for the weak and strong couplings. 
Those values are
consistent with large-$N_c$ and naturalness, but such consistency will not occur
in observables where $h_\pi^1$ contributes.

The $1/N_c$ expansion for hadronic matrix elements, superimposed on
suppressions by factors of $\sin^2 \theta_W$ predicted by the Standard Model,
provides a new benchmark for PV NN couplings. This approach not
only estimates the couplings, it also gives plausible ranges for them based on
$1/N_c$ scaling. The results for the only non-zero measurements of
parity-violating effects in the NN system  are consistent with data. It also naturally predicts a small $h_\pi^1$:
$|h_\pi^1|\lesssim(0.8 \pm 0.3)\times 10^{-7}$ in agreement with 
the bound obtained from 
$^{18}$F experiments~\cite{Barnes:1978,Haxton:1981,Adelberger:1983zz,Evans:1985,Bini:1985,Page:1987ak}, 
$|h_\pi^1| \lesssim 1.3 \times 10^{-7}$. It is also consistent with the first lattice calculation of $h_\pi^1$~\cite{Wasem:2011zz}.
The $1/N_c$ expansion thus explains the otherwise puzzling failure of pion effects to yet manifest themselves in hadronic parity-violation experiments. 
Finally, it also suggests a new $\Delta I=1$ spin-flavor
structure ($U^1_D$) at ${\cal O}(N_c^0 \sin^2 \theta_W)$ should be included in analyses that
examine the subleading piece of the NN force generated by the weak
interaction. 

\vspace*{-4mm}
 
\acknowledgments{We thank Martin Savage and Wick Haxton for useful discussions. We are grateful to Mike Snow for valuable comments and pointing us to pertinent literature. This work was
supported by US Department of Energy grant DE-FG02-93ER40756 (D.~R.~P.). 
The visit of D.S. to Ohio University was supported by TRF-RMUTI under
contract No. TRG5680079 and RMUTI.
C.S. thanks the INPP at Ohio University and the Theory Group 
at ZARM, University of Bremen, for their warm hospitality.}

\end{document}